\documentclass[a4paper,11pt]{article}
\pdfoutput=1 % if your are submitting a pdflatex (i.e. if you have
\usepackage{jinstpub} % for details on the use of the package, please
\usepackage[squaren]{SIunits}
\usepackage{amsmath,amssymb}

\title{Timing performances of a Time-Of-Flight detection system for the FRACAS large acceptance mass spectrometer}

\author[1]{S. Salvador,\note{Corresponding author.}}
\author{E. Barlerin, J. Perronnel and C. Vandamme}
\affiliation{LPC Caen, Normandie Univ, ENSICAEN, UNICAEN, CNRS/IN2P3, 14000 Caen, France}

\emailAdd{salvador@lpccaen.in2p3.fr}

\abstract{In this work, we report on the timing performances of the Time-Of-Flight (TOF) apparatus of the FRACAS large acceptance mass spectrometer designed to measure the fragmentation cross sections of a $^{12}$C beam in hadrontherapy. The TOF system is composed of a Parallel Plate Avalanche Counter and scintillating detectors to create so-called $\Delta$E---TOF maps used in the fragment charge identification process. From tests with an alpha source and a $^{12}$C beam experiment, we measured a coincidence resolving time for the apparatus below 300~ps. Those results should lead in the final design to the identification of all the fragments produced by the interaction of $^{12}$C ions with thin targets for energies up to 400~MeV/nucleon.}


\begin{document}
\maketitle
\flushbottom

\section{Introduction}
\label{sec:intro}

    Carbon ions therapy is being considered to replace proton therapy in the treatment of certain cancerous diseases due to a better ballistic and a higher relative biological effectiveness (RBE). However, nuclear interactions in the tissues between beam particles and nuclei reduce the primary beam intensity at the Bragg peak and create a mixed radiation field through lighter and faster fragments.

    Even though nuclear fragmentation models are included in some Monte Carlo based Treatment Planning Systems (TPS), they still lack sufficient cross section data to accurately reproduce the correct mixed radiation field~\cite{bohlen}.
    Carbon fragmentation cross section experiments have already been conducted to this end by several groups~\cite{deNapoli1,divay,dudouet1,dudouet2}, but those were focused on a lower energy range from 50 to 95~MeV/nucleon. Considering that a carbon therapy treatment can require a beam energy up to 400~MeV/nucleon, fragmentation cross sections up to these energies are of the utmost importance.
    
    In this context, the ARCHADE center under construction in Caen, France, will provide a physics experimental room to study carbon ion therapy with a maximum energy of 400~MeV/nucleon. In this room, $^{12}$C double-differential fragmentation cross sections will be conducted on thin targets (i.e. C, H, O, N and Ca). While particle identification may be achieved using several well-known technique such as the range~\cite{Chulick, Greiner, Amaldi}, Bragg peak amplitude measurements~\cite{Asselineau}, $\Delta$E---E~\cite{Dudouet3} or even pulse-shape discrimination analysis~\cite{Mutterer,LeNeindre}, a time-of-flight spectrometer~\cite{Bass,Banerjee} proved to be the most accurate detection system for ions with kinetic energies higher than 150~MeV/nucleon~\cite{Pleskac,salvador}.

\begin{figure}[!h]
        \begin{center}
            \includegraphics[width=1.0\linewidth]{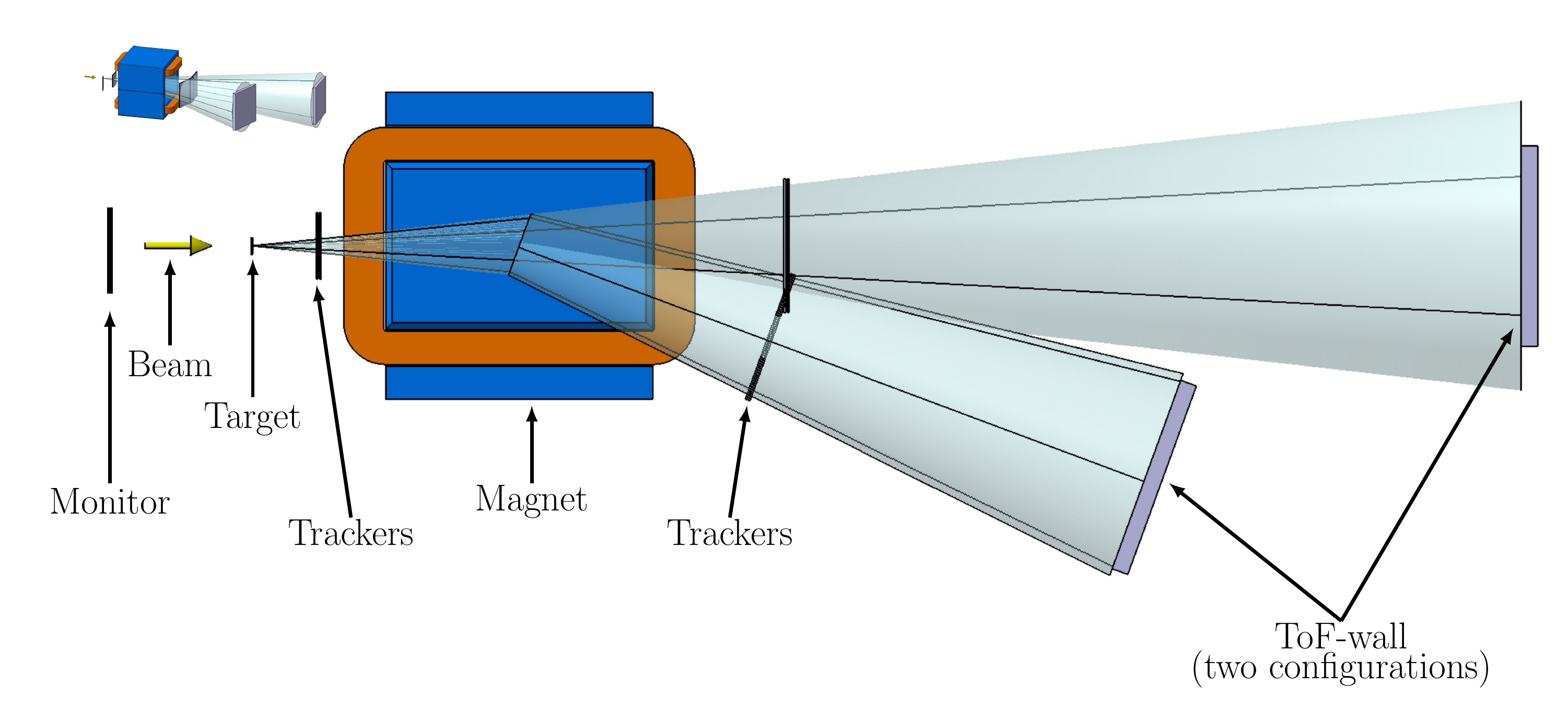}
        \end{center}
        \caption{Basic representation of the large acceptance mass spectrometer FRACAS with the essential detection elements and two different TOF-wall configurations.}
        \label{fig:fracas}
    \end{figure}

    We then designed the FRACAS (FRAgmentation of CArbon and cross Sections) system as a large acceptance mass spectrometer with Time-Of-Flight (TOF) capabilities in order to obtain the missing fragmentation cross sections of a $^{12}$C beam.
    A schematic view of FRACAS is presented in figure~\ref{fig:fracas}. The spectrometer will be made of: i) a TOF-wall involving scintillating detectors to measure the velocity and to extract the charge of the incident fragments; ii) a beam monitor located before the target giving the accurate position and time of arrival of each incident ion; iii) a large acceptance deflecting magnet surrounded up- and down-stream by particle tracking systems for mass identification.

    In this work, we focus mainly on the timing performances of the TOF measurement system where the \emph{start} will be given by a Parallel Plate Avalanche Counter (PPAC), considered as the beam monitor in the final design, in coincidence with a TOF-wall module to obtain the partial deposited energy and the \emph{stop} information. The performances will be evaluated as the FWHM of the coincidence time distributions in terms of Coincidence Resolving Time (CRT).

\section{Methods and experimental set-up}

    The TOF detection system for the FRACAS experiment will be composed of a multi-stage PPAC as the \emph{start} and beam monitor, and a TOF-wall.

    The electrodes of the PPAC time stage were made of 5~$\times$~5~cm$^2$~and 2.5~$\micro$m thick mylar foils where 200~nm~thick Ni/Cr (80\%-20\%) layers were evaporated on both sides. The gas gap was accurately set to 1.6~mm~using epoxy spacers. The negative high voltage was set on the cathode while the signal read-out was taken on the anode. The system was placed inside a small gas vessel where isobutane (iC$_4$H$_{10}$) was flushed at a 0.26~ln/h flux and different low pressures ranging from 20 to 50~mbar.

\begin{figure}[!ht]
        \includegraphics[width=0.475\linewidth]{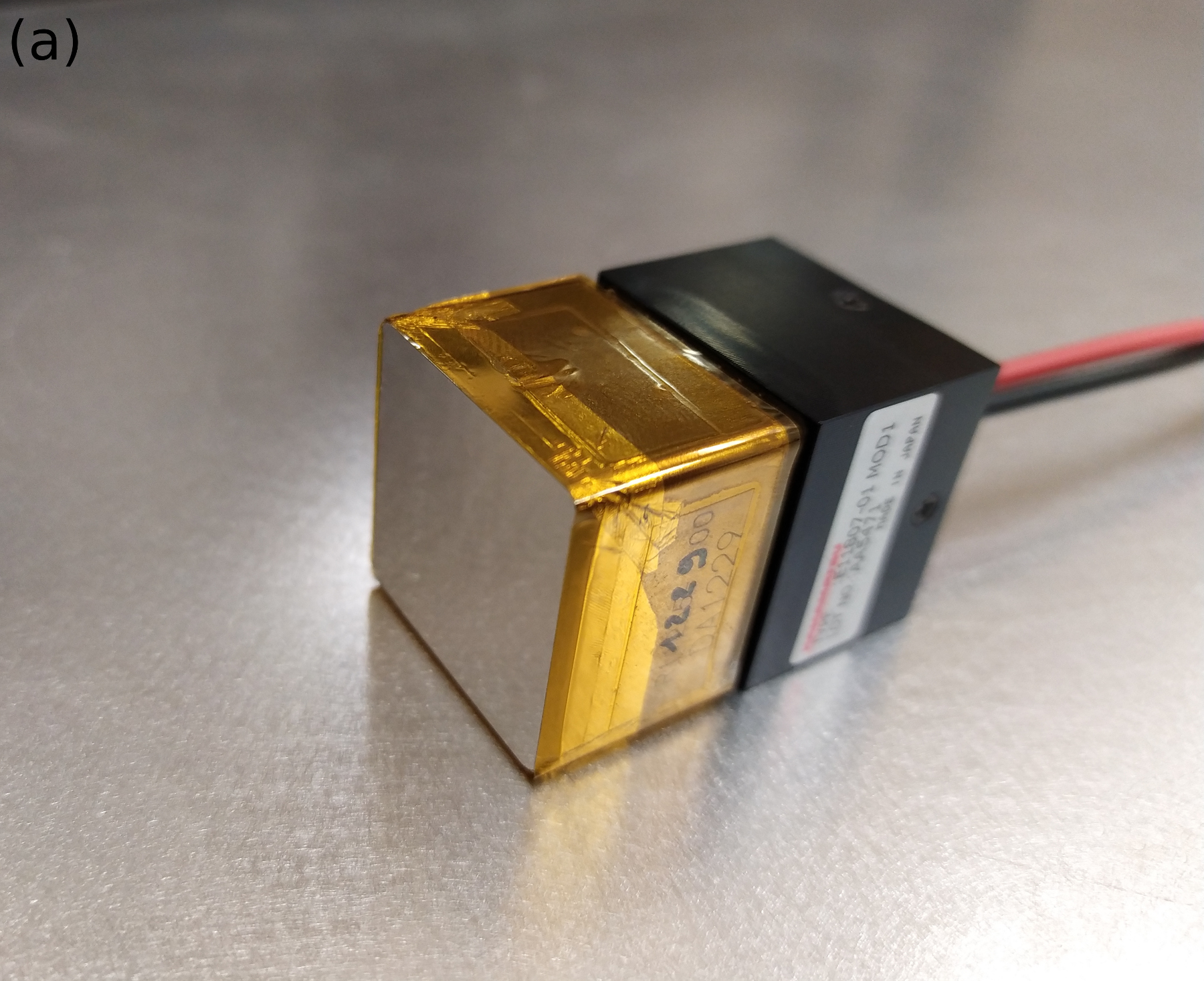}
        \includegraphics[width=0.5\linewidth]{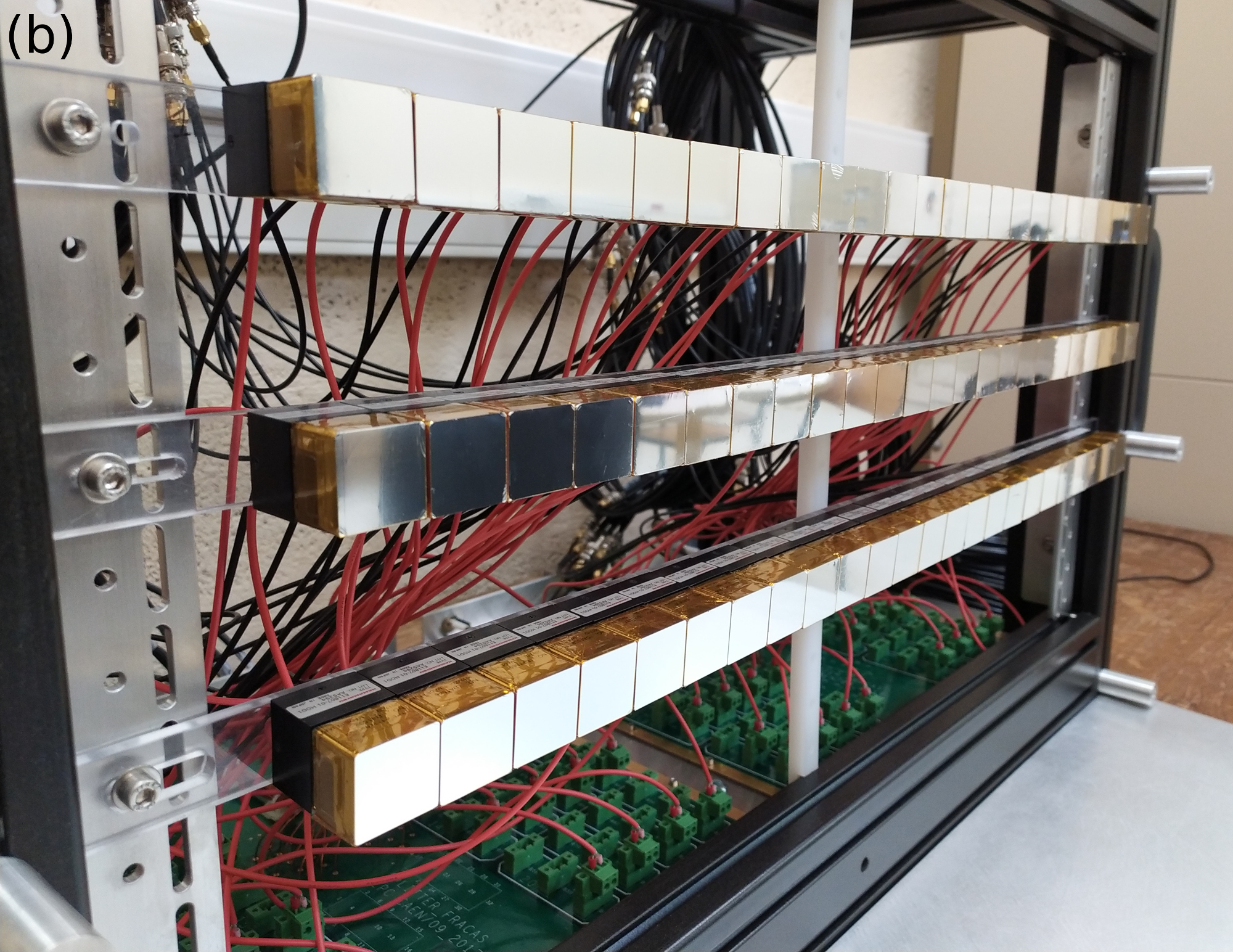}
        \caption{Pictures of (a) a Tof-wall module and (b) 72 modules arranged on the apparatus.}
        \label{fig:pictureModule}
    \end{figure}

    The FRACAS TOF-wall will be a modular system made of 19~$\times$~19 scintillating detector modules that can be arranged in different configurations regarding the experimental program to be done. Each module is composed of an Hamamatsu R11265-200 ultra-bialkali Photomultiplier Tube (PMT) coupled with NUSIL LS-6943 optical polymer to a 25~$\times$~25~$\times$~1.5~mm$^3$~YAP:\,Ce crystal from Crytur Corp. having a rise time $\tau_r$=380~ps and a decay time $\tau_d$=27~ns~\cite{moszynski}. To increase the light collection efficiency while reducing the material budget, the scintillating crystals and PMT windows have been wrapped with 1.5~$\micro$m thick aluminised mylar sheets. As the PMT metallic casing is set at the high voltage, it has been insulated from the other modules with kapton tape (figure~\ref{fig:pictureModule}a). Currently, 72~modules have been mounted and tested on the TOF-wall apparatus (figure~\ref{fig:pictureModule}b).

    The signal of the PPAC was pre-amplified using an in-house built fast pre-amplifier with a 50~$\Omega$ input impedance and a voltage gain of 50. Both signals, from the PMT and the pre-amplified PPAC, were fed to a 10~GS/s Lecroy Waverunner 620Zi oscilloscope. Constant Fraction Discriminators (CFDs) were used for triggering the events and to precisely measure the triggering times by reducing to the minimum the time-walk effect. 

\subsection{Optimization of the CFDs}

\begin{figure}[!ht]
        \includegraphics[width=0.5\linewidth]{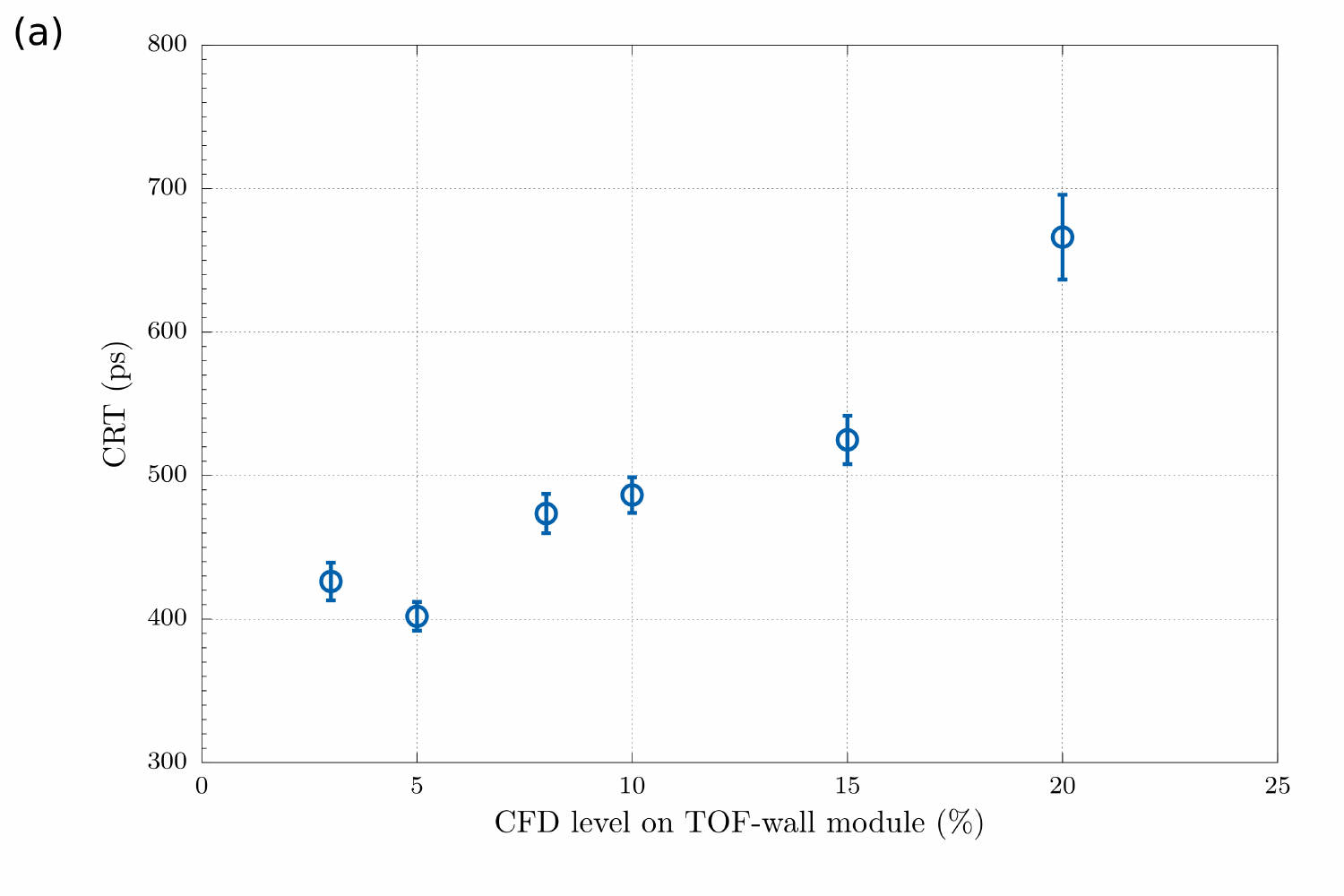}
        \includegraphics[width=0.5\linewidth]{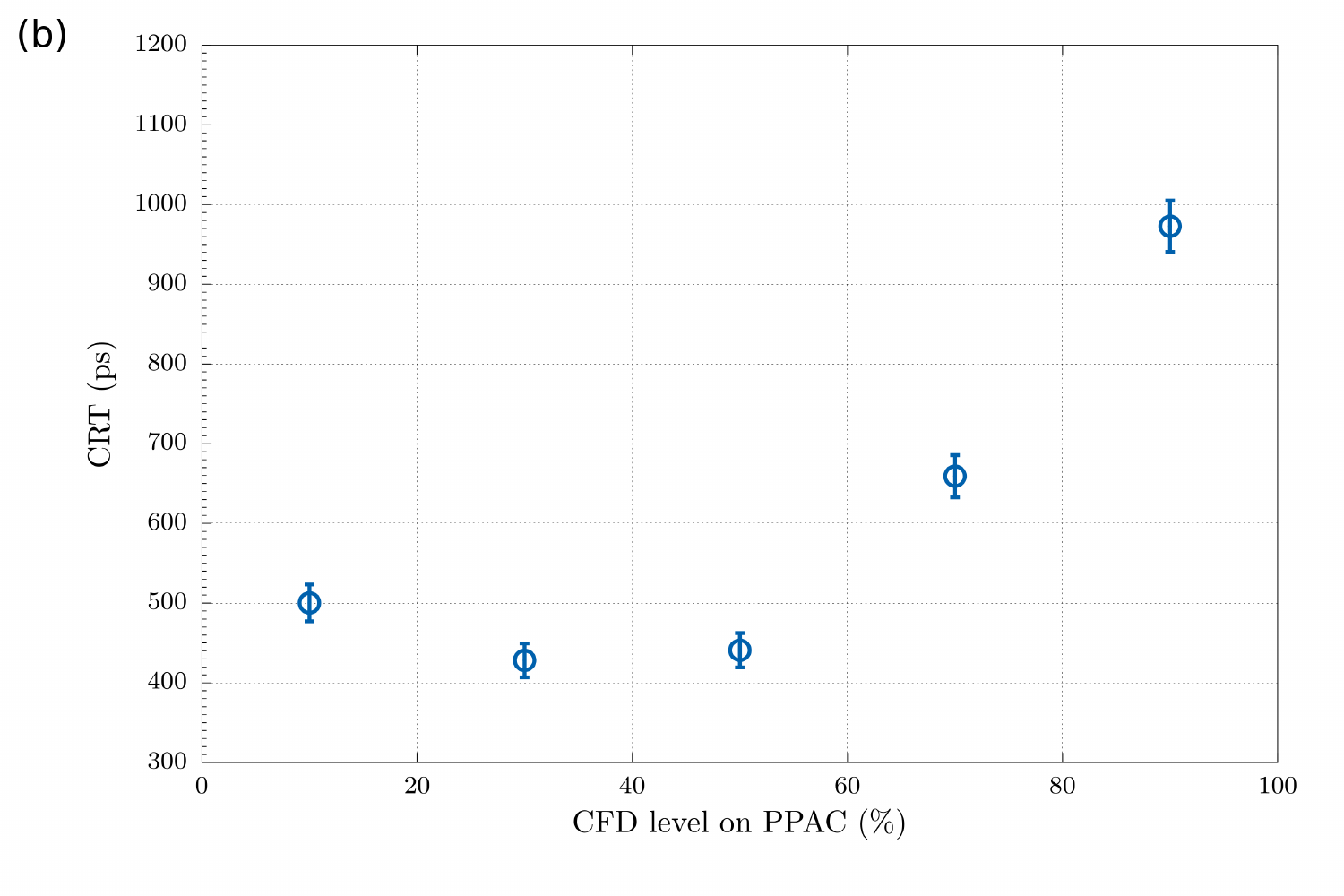}
        \caption{(a) Effect of the CFD level on the CRT for two identical TOF-wall modules in coincidence with a $^{22}$Na source and (b) CRT as a function of the CFD level on the PPAC signals for a gas pressure of 30~mbar.}
        \label{fig:CFD_CRT}
    \end{figure}

    In order to optimize the value of the CFD for the TOF-wall, we put two identical TOF modules in coincidence with a $^{22}$Na source emitting two back-to-back 511 keV $\gamma$-rays. The CRTs were extracted for full energy events in the crystals by applying on both modules an energy window surrounding the full energy peak. Figure~\ref{fig:CFD_CRT}a) shows the well-known CRT behaviour of scintillating detector with respect to the CFD level where it exhibits a minimum around a few percent level threshold. This is widely attributed to the scintillation photon timing and statistics process where the first photons carry more useful time information compared to the later ones which undergo more interactions inside the crystal~\cite{seifert}. The CFD value of 5\% of the signal maximum was then used for the following measurements.
    
    The CFD for the PPAC signals has been optimized during the evaluation of the CRT for the TOF system (described in section~\ref{sec:CRT}). Figure~\ref{fig:CFD_CRT}b) shows the values obtain for a PPAC gas pressure of 30~mbar and a high voltage of -700~V corresponding to a reduced eletric field of 146~kV/(cm$\cdot$bar). The optimal CFD level for the PPAC signals was found to be around 40\%.

\subsection{Table-top CRT evaluation}
    \label{sec:CRT}
\begin{figure}[!ht]
        \begin{center}
            \includegraphics[width=0.6\columnwidth]{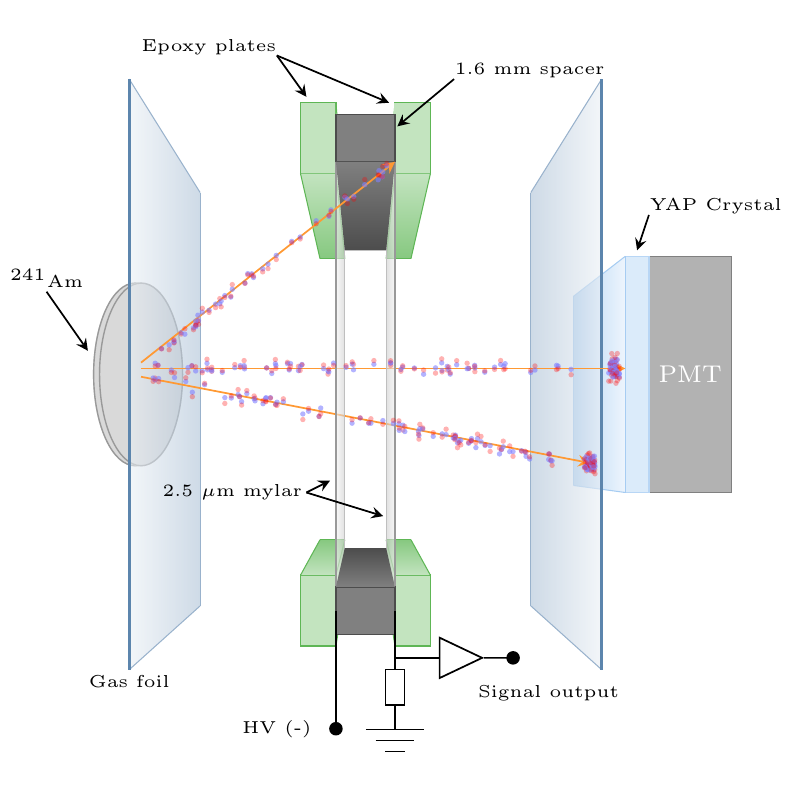}
        \end{center}
        \caption{Sketch of the CRT set-up representing the PPAC and the scintillating detector.}
        \label{fig:sketch}
    \end{figure}

    Figure~\ref{fig:sketch} shows a sketch of the detection set-up to measure the CRT of the TOF system. 
    The distance between the PPAC detector and the TOF-wall module was set to approx.~6cm. The PMT voltage was set to -900~V as the optimal high voltage recommended by the manufacturer. The system was tested using a $^{241}$Am source, emitting mostly $\alpha$ particles of 5.486~MeV, located on the other side of the PPAC gas vessel, opposite to the TOF-wall module. The entire set-up was placed in a vacuum chamber with a pressure of 10$^{-5}$~mbar.

\begin{figure}[!ht]
        \begin{center}
            \includegraphics[width=0.7\columnwidth]{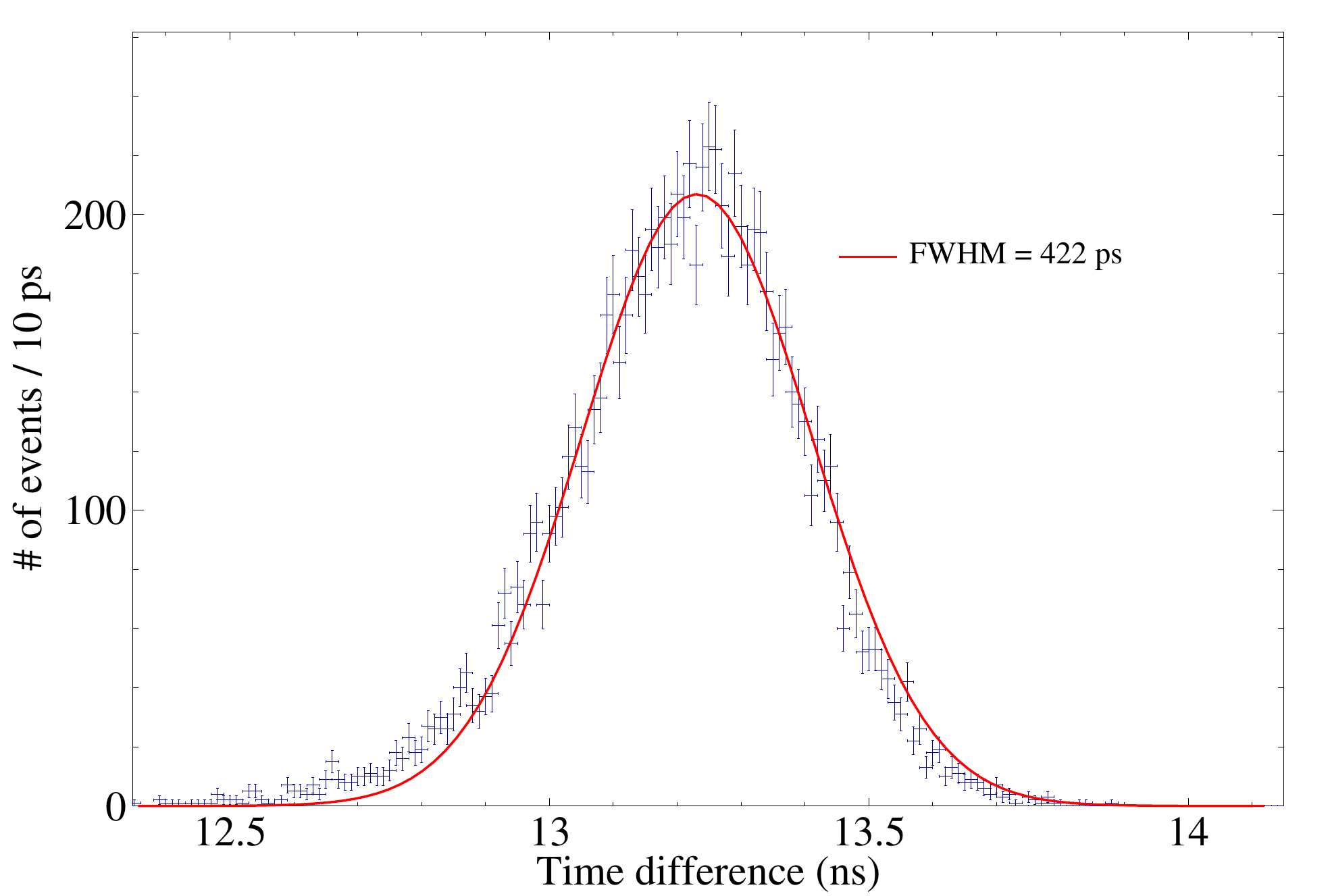}
        \end{center}
        \caption{Example of a time difference spectrum between the PPAC at -700~V and isobutane at 30~mbar, and the PMT signals. The CRT was extracted as the FWHM of the distribution from a Gaussian fit (solid red line).}
        \label{fig:TimeDiff}
    \end{figure}

    Timing measurements were conducted with both signals being triggered within a 30~ns time window using the CFDs. Time difference spectra were then constructed and Gaussian fitted to extract the CRT values as the Full Width at Half Maximum (FWHM) of the distribution. Figure~\ref{fig:TimeDiff} shows an example of a time difference spectrum used to extract the CRT value. CRTs have been evaluated for several gas pressures and applied high voltages.

\subsection{Beam test}

    In order to obtain CRT values for the TOF-wall system under beam conditions, a test was conducted with a 95~MeV/nucleon $^{12}$C beam at GANIL.\footnote{Grand Acc\'el\'erateur National d'Ions Lourds, Caen, France.} However due to experimental constraints, a Micro-Channel Plate (MCP) was used in place of the PPAC as the \emph{start} detector. Ions impinging on the MCP produced secondary electrons in the micro-channels that were collected by an anode. The \emph{stop} of the particles was given by a TOF-wall element at its optimal operating configuration. The entire set-up was installed inside a reaction chamber under 10$^{-7}$~mbar pressure. The distance between the MCP and the TOF module was set to approx.~50~cm.
    
\section{Results}

\subsection{Table-top CRT evaluation}
\begin{figure}[!ht]
        \begin{center}
            \includegraphics[width=0.7\columnwidth]{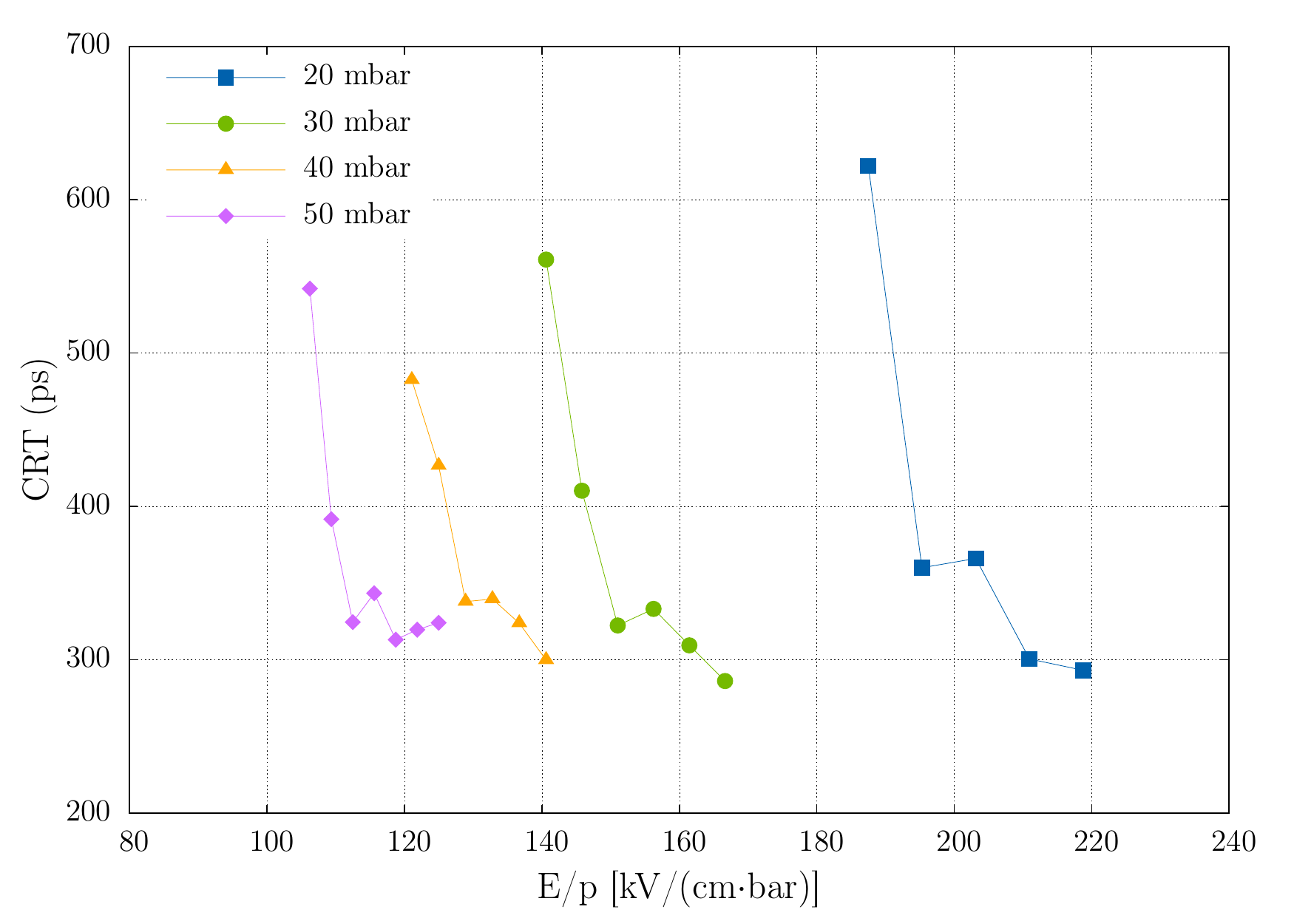}
        \end{center}
        \caption{CRT values measured for several reduced electric fields and gas pressures.}
        \label{fig:TimevsHT}
    \end{figure}

    Figure~\ref{fig:TimevsHT} shows the CRT values obtained for pressures of iC$_4$H$_{10}$ from 20 to 50~mbar and PPAC voltages ranging from $-$600~V to $-$1000~V, expressed as reduced electric field $E/p$. It is clear that increasing the high voltage, i.e. increasing the PPAC gain, leads to improved CRT values, up to the point of the PPAC discharge. Overall, a CRT value below 400~ps was achieved. Values close or below 300~ps can be obtained, but only for gas pressures below 40~mbar and the highest reduced electric fields. In these conditions, the PPAC reached its operating limit and the spark probability began to be non negligeable.

\subsection{Beam test}

\begin{figure}[!ht]
        \begin{center}
            \includegraphics[width=0.7\columnwidth]{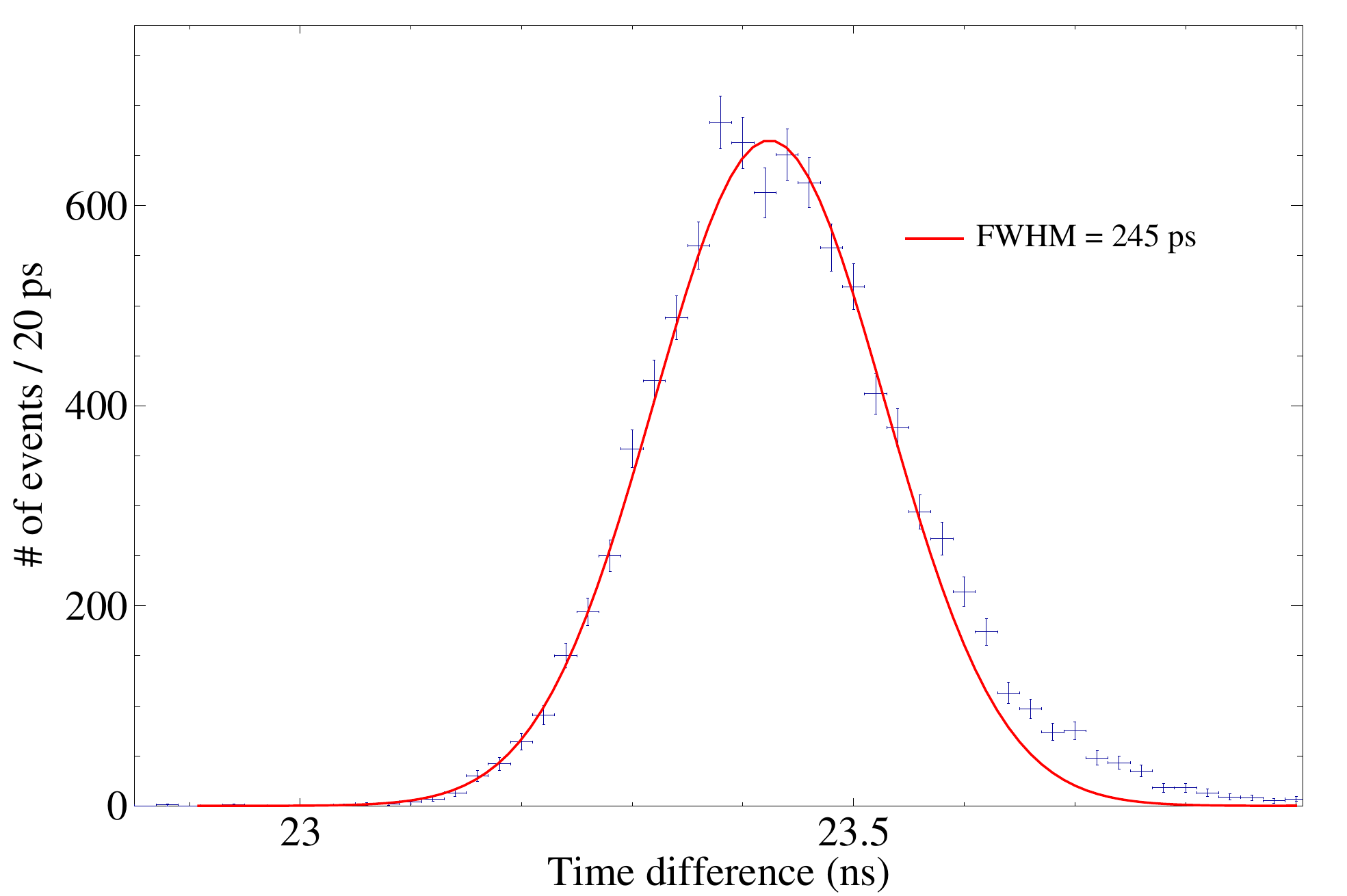}
        \end{center}
        \caption{Time difference spectrum obtained during the $^{12}$C beam test at 95~MeV/nucleon with an MCP in coincidence with a TOF-wall module. The CRT was extracted as the FWHM at 245~ps.}
        \label{fig:Manip0degres_TOF}
    \end{figure}

     The CRT between the MCP and a TOF-wall element was extracted and can be seen on figure~\ref{fig:Manip0degres_TOF}. Even though some beam particles underwent fragmentation processes in the MCP producing lighter particles such as protons and $\alpha$ particles, a 240~ps CRT was extracted. Assuming an intrinsic timing resolution of the MCP of the same order as the PPAC (see for example~\cite{breskin} and~\cite{lienard} for PPAC and MCP time resolutions, respectively), this shows that a CRT value below 300~ps can be obtained in the FRACAS TOF system when the partial deposited energy in the detectors is higher ($\sim$120~MeV deposited in the YAP crystal compared to $\sim$5~MeV for $\alpha$ particles).

\section{Discussion}

    The CRT of a TOF measurement system in a mass spectrometer affects two main aspects in the process of the particle identification. The first is the mass identification resolution, its separation power between different particle isotopes, and the second is the energy resolution. 
    Eqs.~(\ref{eq:massResol}) and~(\ref{eq:enResol}) give the mass and energy resolutions at first order expected in a mass spectrometer, respectively.
    Here, $\rho$ represents the particle trajectory radius in the magnetic field obtained from the magnetic rigidity, $\beta$ and $\gamma$ are the Lorentz factors of the particle. 

\begin{figure}[!ht]
        \includegraphics[width=0.5\columnwidth]{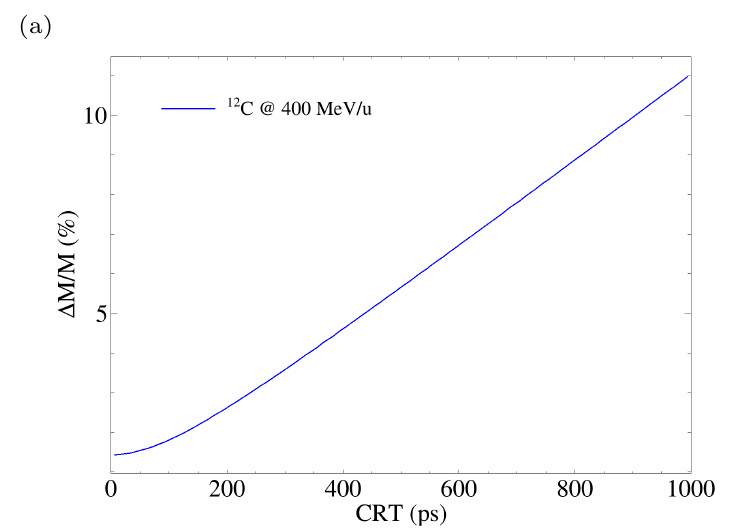}
        \includegraphics[width=0.5\columnwidth]{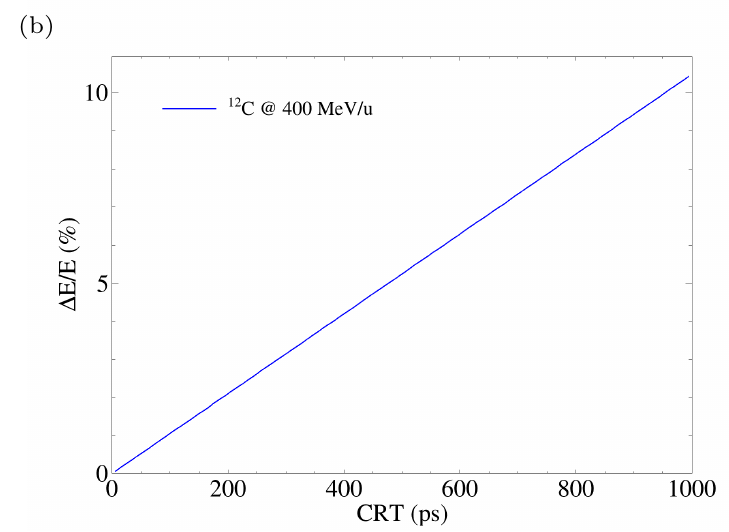}
        \caption{(a) Mass and (b) energy resolution in \% as a function of the CRT of the TOF-wall obtained from eqs.~(\ref{eq:massResol}) and~(\ref{eq:enResol}), respectively. The resolutions have been evaluated for a $^{12}$C particle at 400~MeV/nucleon, a distance of flight of 4~m, a trajectory radius of about 9~m and a TOF around 20~ns.}
        \label{fig:PartResol}
    \end{figure}

    Considering $^{12}$C beam particles at 400~MeV/nucleon, having a radius $\rho\simeq$~9~m with a 4~m flight distance, $\Delta \rho =$~0.15~m (evaluated by Monte Carlo simulations of the FRACAS system in a 0.7~T magnetic field), a TOF value of about 20~ns and a CRT of 300~ps, figures~\ref{fig:PartResol}(a) shows that the mass resolution is theoretically expected to be around 3.2\%. Achieving a good mass separation for the carbon isotopes would require a mass resolution of 8\%, allowing a degradation of the CRT to slightly above 700~ps. Concerning the energy resolution shown on figure~\ref{fig:PartResol}(b), it was evaluated around 3.5\% well under what was achieved using a $\Delta$E---E technique~\cite{dudouet2} or even a similar mass spectrometer~\cite{toppi}. These resolutions can naturally be improved by increasing the distance of flight of the particles but at the expense of the geometrical efficiency.

    \begin{equation}
        \frac{\Delta M}{M} = \sqrt{\left(\frac{\Delta \rho}{\rho}\right)^2 + \left(\gamma^2 \frac{\Delta_{TOF}}{TOF}\right)^2}
        \label{eq:massResol}
    \end{equation}

    \begin{equation}
        \frac{\Delta E}{E} = \frac{\sqrt[3]{\gamma}\beta^2\Delta_{TOF}}{TOF (\gamma - 1)}
        \label{eq:enResol}
    \end{equation}

\section{Conclusion}

In this work, we evaluated the timing performances of the TOF system for the FRACAS apparatus. With a CRT value around 300~ps when using an 5.486~MeV $\alpha$ source and around 250~ps under beam conditions, we expect a mass resolution and a kinetic energy resolution better than 5\%. These results are sufficient to obtain the double differential cross sections of carbon ions required for hadrontherapy. However, further data must be taken to confirm those good results under beam conditions when using the PPAC detection system as the TOF \emph{start}. Experiments are already scheduled in a near futur with proton and $\alpha$ beams to evaluate those performances.

\end{document}